\documentstyle[12pt]{article}

\renewcommand{\baselinestretch}{1.4}

\def\beq{\begin{equation}}
\def\eeq{\end{equation}}
\def\bea{\begin{eqnarray}}
\def\nn{\nonumber}
\def\eea{\end{eqnarray}}
\def\ds{\displaystyle}

\def\ms{\medskip}
\def\bs{\bigskip}

\def\wh{\widehat}
\def\ni{\noindent}

\def\req#1{(\ref{#1})}
\def\Tr{{\rm Tr}\ }

\begin{document}



\vspace*{3mm}

\begin{center}

{\LARGE \bf
Schwinger-Dyson Equations in $2D$ Induced
Gravity in Covariant Gauges}

\renewcommand\baselinestretch{0.8}

\vspace*{0.5cm}
{\rm {\sc E. Elizalde}\footnote{E-mail:
 eli@zeta.ecm.ub.es.
Address june-september 1994:  Department of Physics, Faculty of Science,
 Hiroshima University, Higashi-Hiroshima 724, Japan.
} \\
 Center for Advanced Studies, CSIC, Cam{\'\i} de Santa B{\`a}rbara,
17300 Blanes, \\
and Department ECM and IFAE, Faculty of Physics, University of
Barcelona, \\ Diagonal 647, 08028 Barcelona, Catalonia, Spain, \\
\vspace{0.2cm}
{\sc S.D. Odintsov}}\footnote{
On leave of absence from Dept. of Mathematics and Physics,
Pedagogical Institute, 634041 Tomsk, Russia.
E-mail: odintsov@ebubecm1.bitnet} \\
Department ECM, Faculty of Physics, University of Barcelona, \\
Diagonal 647, 08028 Barcelona, Catalonia, Spain, \\
\vspace{0.2cm}
and \\
{\sc Yu.I. Shil'nov} \\
Department of Theoretical Physics, Kharkov State University, \\
Svobody sq. 4, Kharkov 310077, Ukraine.

\ms

\renewcommand\baselinestretch{1.4}

\vspace{5mm}

{\bf Abstract}

\end{center}

We formulate the Schwinger-Dyson equations in the ladder
approximation
for $2D$
induced quantum gravity with fermions using covariant gauges of
harmonic
type. It is shown that these equations can be formulated
consistently in
a gauge of Landau type (for negative cosmological constant).
 A numerical analysis  of the equations hints towards the
possibility
 of chiral symmetry breaking, depending on the value of the
coupling
constant.

\newpage

A very useful tool for the study of dynamical symmetry breaking and
dynamical mass generation in quantum field theory are the
Schwinger-Dyson equations. The standard approach, which consists in
working with this infinite set of integral equations has been
developed
in the pioneering works \cite{1,2}, where the example of quantum
electrodynamics was considered. By investigating the truncated
version of the Schwinger-Dyson equations (the ladder
approximation), the
possibility of chiral symmetry breaking and dynamical fermion mass
generation in QED could be demonstrated \cite{1,2} (for a review,
see the
proceedings \cite{3}).  The critical coupling constant in the
Landau
gauge has been found also. However, if dynamical fermion mass
generation
is taking place, then the non-perturbative Ward-Takahashi identity
of
QED is not satisfied. As a result, the dynamical fermion mass and
the
critical coupling constant in QED are very much gauge-dependent
(for a
recent discussion in an arbitrary covariant gauge, see \cite{4,5}
and
references therein).

If one is interested in further developing the Schwinger-Dyson
formalism
then more complicated models to study such equations, like those
of
quantum gravity, are to be considered. An example of an
investigation of this kind for the case of 4D Einstein gravity coupled
to fermions on
a flat background has been presented in ref. \cite{6}, where by
means of
numerical estimations the possibility of chiral symmetry breaking
has
again been shown.

In the  present work we will  study the Schwinger-Dyson
equations in 2D quantum gravity \cite{7} with fermions, on a flat
background. The covariant gauge with two gauge parameters will be
chosen
and chiral symmetry breaking in 2D quantum gravity will be
investigated
numerically.

The action of the theory under discussion is given by
\bea
S&=&S_g+S_f, \nn \\
S_g&=&\ds -{1 \over 2 \gamma} \int d^2x \ \sqrt{-g}
\left( R {1 \over \Box} R + \Lambda \right) , \nn \\
S_f&=&\ds \int d^2x \ \sqrt{-g} \ i \ \bar\Psi \gamma^{\mu} D_{\mu}
\Psi,
\label{1}
\eea
where $R$ is the two-dimensional curvature, $\Psi$ the 2D spinor,
$\Lambda$ the cosmological constant, and $D_{\mu}= \partial_\mu -
(i/4)
\omega^{bc}_{ \ \mu} \, \sigma_{bc} $ is the
2D covariant derivative for spinors, where $\sigma_{ab} = (i/2)
[\gamma_a, \gamma_b]$, $\omega^{bc}_{ \ \mu}$ is
spin-connection, and $e^a_{\ \mu}$ will denote the {\it vierbein}.

 In the standard approach
to string theory \cite{7}, one can start from a pure-matter theory
(as
given by $S_f$) in an external gravitational field, integrate then
over
spinors (this is easy to do in the conformal gauge) and get finally as
a
result 2D induced gravity, $S_g$, where $\gamma$ is then specified.

Here we are going to employ a more traditional approach, in which we
will
start from the theory (\ref{1}) and use the background field method
on
the flat background
\beq g_{\mu \nu}=\eta_{\mu \nu} + h_{\mu\nu}, \ \ \
 e_{a \mu}=\eta_{a \mu} + \frac{1}{2} h_{a\mu}.
 \label{2}
\eeq
Hence, $\gamma$ in (\ref{1}) is some given constant and we do not
integrate over spinors. Expanding $S_f$ on the flat background and
working in the momentum representation, one easily finds that the
interaction Lagrangian has the following form
\beq
L_{\rm int}={1 \over 4} \bar\Psi (p')
\left[ 2 \wh{p} \eta_{\mu \nu} -2 \gamma_{(\mu} p_{\nu)}
+ \left( \wh{k} \eta_{\mu \nu} - \gamma_{(\mu} k_{\nu)} \right)
\right] \Psi (p)  h^{\mu\nu} (k),
\label{3}
\eeq
what corresponds to the fermion-graviton vertex
\bea
\Gamma_{\mu\nu} (p,k)&=& \frac{1}{4} (2p+k)^\lambda \gamma^\sigma
I_{\lambda\sigma\mu\nu}, \nn \\
I_{\lambda\sigma\mu\nu}&=& \frac{1}{4} \left( 2\eta_{\lambda\sigma}
\eta_{\mu \nu} -\eta_{\lambda\mu} \eta_{\nu \sigma} - \eta_{\lambda
\nu} \eta_{\sigma \mu} \right). \label{4}
\eea
The gauge-fixing action will be choosen in the following form
\beq
S_{gf} = \ds -{1 \over 2 \gamma} \int d^2x \ \sqrt{-g} \,
\frac{1}{\alpha} \left( \nabla_\mu h^\mu_{\ \rho} - \beta
\nabla_\rho h \right) \left( \nabla_\nu h^{\nu \rho} - \beta
\nabla^\rho h \right), \label{5}
\eeq
where $\alpha$ and $\beta$ are the gauge parameters. (For a
discussion of 2D induced gravity in a covariant gauge of the
harmonic type, see also \cite{8,9}).

The quadratic part of the total action $S=S_g + S_{gf}$ on a flat
background is found to be
\beq
S^{(2)} = \ds -{1 \over 2 \gamma} \int d^2x \ h^{\mu\nu}
H_{\mu\nu\rho\sigma} h^{\rho\sigma},
\label{6}
\eeq
where
\bea
H_{\mu\nu\rho\sigma} &=& \frac{\nabla_\mu\nabla_\nu
\nabla_\rho\nabla_\sigma}{\Box} + \frac{1}{2}\xi_1 \left(
\eta_{\rho\sigma} \nabla_\mu\nabla_\nu + \eta_{\mu\nu}
\nabla_\rho\nabla_\sigma \right) + \xi_2 \eta_{\mu\nu}
\eta_{\rho\sigma} \Box \nn \\
&& + \frac{1}{4}\xi_3 \left( \eta_{\mu\sigma} \nabla_\nu\nabla_\rho
+ \eta_{\mu\rho} \nabla_\nu \nabla_\sigma  + \eta_{\nu\sigma}
\nabla_\mu\nabla_\rho + \eta_{\nu\rho} \nabla_\mu \nabla_\sigma
\right) + \frac{\Lambda}{4}  I_{\mu\nu\rho\sigma},
\eea
being $\xi_1 =2(\beta /\alpha -1)$, $\xi_2 = 1- \beta^2/\alpha$ and
$\xi_3 =-1/\alpha$. The graviton propagator is given by the inverse
operator $H_{\mu\nu\rho\sigma}^{-1}$ and can be found using the
algorithm of refs. \cite{10}, which yield
\bea
&&G_{\mu\nu\rho\sigma} (k) = -\gamma H_{\mu\nu\rho\sigma}^{-1} (k)
= -\frac{4}{\Lambda} L_{\mu\nu\rho\sigma} +\frac{1}{\alpha -(2\beta
-1)^2} \left( \frac{ [\alpha + \beta (1-2\beta)]^2}{(\beta -1)^2
(k^2-m^2)} - \frac{\alpha}{k^2} \right) P_{\mu\nu\rho\sigma} \nn \\
&& +\frac{2\alpha }{k^2- m^2}
M_{\mu\nu\rho\sigma}
+\frac{1}{\alpha -(2\beta -1)^2} \left( \frac{ (2\beta -1) [\alpha
+ \beta (1-2\beta)]}{(\beta -1) (k^2-m^2)} - \frac{\alpha}{k^2}
\right) (L_{\mu\nu} P_{\rho\sigma} +L_{\rho\sigma}P_{\mu\nu} ) \nn
\\ && \hspace{15mm} +\left[ \frac{4}{\Lambda} +\frac{1}{\alpha -(2\beta
-1)^2}
\left( \frac{ (1-2\beta)^2}{(k^2-m^2)} - \frac{\alpha}{k^2} \right)
\right] L_{\mu\nu} L_{\rho\sigma}.
\label{7}
\eea
Here
\beq
 L_{\mu\nu} =\eta_{\mu\nu} - \frac{ k_\mu k_\nu }{k^2} , \ \ \
L_{\mu\nu\rho\sigma} = \frac{1}{2} \left(L_{\mu\rho} L_{\nu\sigma}
+L_{\mu\sigma}L_{\nu\rho} \right), \ \ \  P_{\mu\nu} = \frac{ k_\mu
k_\nu }{k^2},
\label{8} \eeq
\[ P_{\mu\nu\rho\sigma} = \frac{ k_\mu k_\nu k_\rho k_\sigma
}{k^4}, \ \ \ M_{\mu\nu\rho\sigma} = \frac{1}{2} \left(L_{\mu\rho}
P_{\nu\sigma} +L_{\mu\sigma}P_{\nu\rho} + L_{\nu\rho} P_{\mu\sigma}
+L_{\nu\sigma}P_{\mu\rho} \right), \ \ \  m^2 = \frac{ \alpha
\Lambda}{2}.  \]
The exact spinor propagator has the following form
\beq
S^{-1} (p) = {\cal A} (p) \wh{p} - {\cal B} (p^2),
\label{9}
\eeq
where $\cal A$ and $\cal B$ are some unknown functions.  Now we have at
hand all the Feynman diagram elements: the exact spinor propagator, the
free spinor porpagator, $S_0^{-1} (p) =  \wh{p}$, the free graviton
propagator (\ref{7}) and the vertex (\ref{4}).

The effective potential for the composite fields
\cite{11} in the ladder approximation \cite{1,2} can be written as
\beq
V_{\rm eff}= -i \ {\rm Sp} \left( \ln S_0^{-1}S - S_0^{-1} S +1 \right)
+ V_2, \label{10}\eeq
where $V_2$ corresponds to the two-particle irreducible vacuum
diagram, which follows from the vertex
\beq
V_2=-{i \over 2} \int
{d^2 p \over (4\pi)^2} \int {d^2 q \over (4\pi)^2}
\Tr \left[ \Gamma(p-q,q) S(q) \Gamma(q-p,p) G(p) \right] .
\label{11}
\eeq

The Schwinger-Dyson equations (in the ladder approximation)
correspond to the minimum of the potential (\ref{10})
\beq
S^{-1} (p) - S_0^{-1} (p) =-i \int {d^2 q \over (4\pi)^2}
\Gamma_{\mu\nu}(q,p-q) S(q) \Gamma_{\rho\sigma}(p,q-p)
G^{\mu\nu\rho\sigma}(p-q).
\label{12}
\eeq
Using the explicit form of the spinor and graviton propagators
(\ref{9}), (\ref{7}), and the vertex (\ref{4}), one can get
---after performing Wick's rotation and the angular integration (we
drop the details of these straightforward but very tedious
calculations)
\[
V_{\rm eff}=-{N_f M^2 \over 8 \pi}\left\{
\int_0^1 dx \left[ \ln\left( A^2(x) + { B^2(x) \over x } \right)
-2 { A(x)(A(x)-1)x+B^2(x) \over x A^2(x) + B^2(x) } \right] \right.
\]
\beq
\left. +g \int_0^1 {dx \over x A^2(x)+B^2(x) }
\int_0^1 {dy \over y A^2(y)+B^2(y) }
[ A(x)A(y)K_A(x,y)+B(x)B(y)K_B(x,y) ] \right\} ,
\label{13}\eeq
where $N_f$ is the dimension of the fermion representation, $M$
the
momentum cutoff, $x=p^2/M^2$,
$y=q^2/M^2$, $A(x)={\cal A}(p^2)$, $B(x)={\cal B}(p^2)/M$,  and
$g=\gamma /( 64 \pi)$ and $l=\Lambda /( 2M^2 )$. The explicit
expressions for $K_A$ and $K_B$ are very complicated for arbitrary
$\alpha$ and $\beta$. Moreover, if $\alpha \neq 0$ the
Schwinger-Dyson equations contain the infrared divergences caused
by the graviton zero momentum. Let us give some examples of $K_A$
and $K_B$ for different choices of the gauge parameters.
\ms

\ni 1. Gauge with $\alpha$ arbitrary, $\beta = 1/2$.
\[
K_A(x,y)=\ds\frac{1}{2} \left\{ (1-3\alpha )(x+y) + { \alpha [7
(x^2+y^2) +10xy+3\alpha l (x+y)] +4(x+y)(x-y)^2/l  \over
\sqrt{(x+y+\alpha l)^2-4xy} } \right. \] \[ \hspace{3cm} \left. -
{x^2+y^2 +6xy +4 (x+y)(x-y)^2 /l \over |x-y|} \right\}, \]
\beq K_B(x,y)=\ds 5\alpha -1 -{ 5\alpha [ 2(x+y)+ \alpha l] + 4 (x-
y)^2 /l \over
\sqrt{(x+y+\alpha l)^2-4xy} }  + {2 (x+y) + 4 (x-y)^2 /l \over
|x- y|}. \label{14}
\eeq
Here $\Lambda > 0$  and $l=\Lambda /( 2M^2 )$, and in this gauge
one finds  infrared
divergences in the Schwinger-Dyson equations (i.e., at the lower
limit of the integrals in (\ref{13})). \ms

\ni 2. Let us now consider a gauge of Landau type ($\alpha =0$,
$\beta$ arbitrary), in which the Schwinger-Dyson equations do not
contain infrared divergences. There one finds
\[
K_A(x,y)=\ds\frac{1}{2} \left\{ \frac{\beta^2}{(\beta -1)^2}
(x+y) - \left[ { \beta [4 \beta xy + 2(2\beta -1) (x-y)^2 +\beta
(x+y) (x+y + \mu^2)] \over
(\beta -1)^2} \right. \right. \] \[ + \left. \frac{4(4\beta -1)}{l
(2\beta -1)} (x+y ) (x-y)^2
 + \left( \frac{4(\beta -1)}{l(2\beta -1)}
\right)^2 (x-y)^4 \right]  \left[ (x^2+y^2+\mu^2)^2-4xy \right]^{-
1/2} \] \[  \hspace{3cm} \left. +  \frac{8\beta }{l(2\beta -1)}
(x+y) |x-y|
 + \left(
\frac{4(\beta -1)}{l(2\beta -1)} \right)^2 |x-y|^3 \right\}, \]
\[ K_B(x,y)=\ds \left[ \frac{\beta^2}{(\beta -1)^2} [2(x+y) +\mu^2]
+ \frac{4 (x-y)^2}{l(2\beta -1)} \right] \left[ (x^2+y^2+\mu^2)^2-4xy
\right]^{-1/2} \] \beq \hspace{3cm} -\frac{4|x-y|}{l(2\beta -1)}
-\frac{\beta^2}{(\beta -1)^2}. \label{15}
\eeq
Here $\Lambda < 0$, $l=-\Lambda /( 2M^2 )$ and $\mu^2 = (4/l)
[(2\beta -1)/(\beta -1)]^2$. Notice that for $\alpha =0$, $\beta
=1/2$, the theory contains again infrared divergences, because in
this case
\bea
K_A(x,y)&=&\ds\frac{1}{2} \left( x+y - \frac{x^2+y^2+6xy}{|x-y|}
\right), \nn \\
K_B(x,y)&=& 2 \frac{x+y}{|x-y|} -1. \label{16}
\eea
Observe also that, in principle, one expect to find more
complicated covariant gauges which are free of infrared problems in
the region where the cosmological constant is positive.

Starting from eqs. (\ref{12}) and integrating over the angles one
can show that the functions $A$ and $B$ must obey integral
equations of the following form
\bea
A(x)&=&\ds 1+g\int_0^1 dy {A(y) \over yA^2(y)+B^2(y) }{1 \over x}
K_A(x,y), \nn \\
B(x)&=&\ds g\int_0^1 dy {B(y) \over yA^2(y)+B^2(y) } K_B(x,y) .
\label{17}
\eea
It is not possible to solve these equations analytically. (We will
discuss here the case of the physical Landau-type gauge (\ref{15})
only, where no IR divergences appear in the theory, in order to
avoid the introduction of any IR cutoff).
We present the result of a numerical calculation, obtained  by
using an iterative procedure (in close analogy with \cite{6}). We
consider two types of trial functions
\[
\begin{array}{lll}
\mbox{(a)}&A^0(x)=c_1,&B^0(x)=0, \\
\mbox{(b)}&A^0(x)=c_1,&B^0(x)=c_2,
\end{array}
\]
where $c_1$ and $c_2$ are some constants between 0 and 1. We will
also fix the values of $g$, $l$ and $\beta$. The functions
$A^0(x)$
and $B^0(x)$ can then be taken as the starting point of a
self-consistent iterative
calculation of the form
\bea
A^{i+1}(x)&=&\ds 1+g\int_0^1 dy {A^i(y) \over
y{A^i}^2(y)+{B^i}^2(y) }
{1 \over x} K_A(x,y), \nn \\
B^{i+1}(x)&=&\ds g\int_0^1 dy {B^i(y) \over y{A^i}^2(y)+{B^i}^2(y)
} K_B(x,y) . \label{18}
 \eea
The sequences formed by the $\{ A^i(x) \}$ and $\{ B^i(x) \}$ are
expected
to converge towards the functions $A(x)$ and $B(x)$, respectively,
which are the sought for  solutions of \req{17}. In practice one
can judge the
degree of convergence of these series by the smallness of the
squared
norms of the differences $A^{i+1}-A^i$ and $B^{i+1}-B^i$, which we
 set at $10^{-4}-10^{-6}$ in our calculation.
If, for the given $g$ and $l$, there are solutions of both types,
(a) and (b), only the most stable of both by $V_{\rm eff}$ \req{9}
is to be  chosen as the one corresponding to the true vacuum.

We have executed this algorithm to solve \req{18}, starting from
the trial functions (a) and (b), for fixed $l=4$, $\beta = 1/3$
and varying $g$.
For very small $g$'s, both types lead to curves close to $A(x)=1$,
$B(x)=0$, i.e. the chiral symmetric solution, as was to be
expected. As $g$ increases, the value of $V_{eff}$ for the chiral
solution of symmetric type (a) appears to be slightly higher than the
corresponding one for the non-symmetric solution. In particular, for
$g=0.1$ (see Fig. 1), the chiral symmetric solution is the preferred
one. For $g=0.2$ or $g=0.3$ (see Fig. 1 again, where typical curves for
$A$ and
$B$ are presented), one can see that the chiral non-symmetric solutions
are preferable. Hence, we see clearly that the Schwinger-Dyson equations
for 2D gravity with fermions may have chiral symmetry breaking regimes
in the covariant gauges.

We will now say a few words about the regime of the Schwinger-Dyson
equations corresponding to a theory with positive cosmological constant.
In this case, when working in the covariant gauges under discussion, one
encounters problems related with infrared divergences. We will use the
conformal gauge
\beq
g_{\mu\nu} = e^\varphi \eta_{\mu\nu}.
\label{19}
\eeq
In this gauge one finds the Schwinger-Dyson equations (\ref{12}) and
(\ref{17}) with the functions $K_A$ and $K_B$ \cite{12} given by
\bea
K_A(x,y)&=&\ds-{ 4xy+(x+y)(x+y+l-\sqrt{(x+y+l)^2-4xy}) \over
2\sqrt{(x+y+l)^2-4xy} }, \nn \\
K_B(x,y)&=&\ds{ 2(x+y)+l-\sqrt{(x+y+l)^2-4xy} \over
\sqrt{(x+y+l)^2-4xy} }.
\label{20}
\eea
Numerical solutions of the corresponding Schwinger-Dyson equations can
be obtained as above (see ref. \cite{12}). Typical curves for $l=0.5$
and
varying $g$ are shown in Fig. 2. Here we observe again the possibility
of chiral symmetry breaking.

Summing up, we have studied the Schwinger-Dyson equations corresponding
to 2D gravity coupled with fermions in a covariant (harmonic) gauge.
Numerical analysis of the equations show clearly the possibility of
chiral symmetry breaking in the region with negative cosmological
constant, where the Schwinger-Dyson equations can be consistently
formulated in a gauge of Landau type and no infrared divergences appear.
In the region of positive cosmological constant, the analysis done in
the conformal gauge shows as well the possibility of chiral symmetry
breaking. The results of the numerical analysis of the solutions (and
 the Schwinger-Dyson equations themselves) are certainly gauge
dependent. Currently there is no way to solve such a drawback of the
Schwinger-Dyson equations, even in the case of renormalizable theories,
as QED, where the Ward-Takahashi identities have a quite simple form.
We have nothing to add here that can help to resolve this
general problem of gauge dependence \cite{4,5}. Our purpose has been
simply to show that chiral symmetry breaking is indeed possible in 2D
gravity theories with fermions in different gauges. That this is
actually the case has been realized by means of a rather straightforward
numerical analysis of the corresponding Schwinger-Dyson equations.
\vskip5mm

\ni{\Large \bf Acknowledgements}

S.D.O. would like to thank T. Muta for helpful discussions on related
problems and the members of the Department E.C.M., Barcelona University,
for the kind hospitality. We are very grateful to  A. Romeo for his help
with the numerical calculations. This work has
been supported in part by CIRIT (Generalitat de Catalunya), by
DGICYT (Spain), project no. PB90-0022, and by the Russian Foundation for
Fundamental Research, project no. 94-02-03234.

\newpage

\ni{\Large \bf Figure captions}
\bs

\ni{\bf Fig. 1}.
Plot of the functions $A$ and $B$  obtained as the (a)-type
solutions for $g=0.1$ and (b)-type solutions for
$g=0.2$ and $g=0.3$ keeping $l=4$ fixed.
\bs

\ni{\bf Fig. 2}.
Plot of the functions $A$ and $B$  obtained as the (a)-type and (b)-type
solutions for
$g=$0.1, 0.2 and 0.25, keeping $l=0.5$ fixed. Notice how $B$
deviates
more and more from the $g=0$ solution ($B(x)=0$) as $g$ increases.
Although not shown in the figure, the curve keeps going up for
larger
values of $g$.
\end{document}